\documentclass[11pt]{article} 
\pdfoutput=1
\usepackage{amsthm,amsmath,amssymb,xcolor,sectsty,latexsym,mathtools,hyperref,mathrsfs,slashed,bm,epsfig,dcolumn,cite,feynmp,setspace}
\numberwithin{equation}{section}
\usepackage[textwidth=6.5in,top=1.2in,bottom=1.2in]{geometry}
\newcommand{\beq}{\begin{equation}}
\newcommand{\eeq}{\end{equation}}
\newcommand{\bea}{\begin{eqnarray}}
\newcommand{\eea}{\end{eqnarray}}
\newcommand{\bef}{\begin{figure}}
\newcommand{\eef}{\end{figure}}

\newcommand{\mpl}{M_{\rm Pl}}

\newcommand{\gam}{\gamma}
\newcommand{\Veff}{V_{\rm eff}}

\begin{document}

\thispagestyle{empty}
\begin{titlepage}
\nopagebreak

\title{  \begin{center}\bf Quantum Fine-Tuning in Stringy Quintessence Models \end{center} }

\vfill
\author{Mark P.~Hertzberg$^{1,2}$\footnote{mark.hertzberg@tufts.edu}, ~ McCullen Sandora$^{1,2}$\footnote{mccullen.sandora@tufts.edu}, Mark Trodden$^{2}$\footnote{trodden@physics.upenn.edu}}
\date{ }

\maketitle

\begin{center}
	\vspace{-0.7cm}
	{\it  $^{1}$Institute of Cosmology, Department of Physics and Astronomy}\\
	{\it  Tufts University, Medford, MA 02155, USA}\\
	{\it $^{2}$Center for Particle Cosmology, Department of Physics and Astronomy}\\
	{\it University of Pennsylvania, Philadelphia, PA 19104, USA}
	
\end{center}
%\vfill
\bigskip
\begin{abstract}
We investigate the extent to which quintessence models for dark energy are fine-tuned in the context of recent swampland conjectures. In particular, the issue is whether there is a double fine-tuning in which both $V$ and $|\nabla V|$ are fine-tuned, or whether there is only a single fine-tuning due to the relation $|\nabla V|\sim V/\mpl$ arising naturally. We find that indeed this relation arises naturally in simple string compactifications for some scalars, such as the dilaton and volume modulus, when treated classically. However, we find that quantum effects can spoil this natural relation, unless the scalar is conformally coupled to the matter sector. Furthermore, it is well known that such conformal couplings are generically ruled out by fifth force tests. To avoid these fifth forces, an interesting proposal is to assume the scalar (quintessence) only couples to the hidden sector. However, we then find quantum corrections to $V$ from visible sector Standard Model particles generically spoil the relation. A possible way out of all these problems is to have the scalar conformally coupled to a dark sector that is an exact copy of the Standard Model. This ensures the relation $|\nabla V|\sim V/\mpl$ is maintained naturally even when matter particles run in the loop. However, we find that quantum corrections from quintessons or gravitons in the loop spoil the relation if the effective theory has a cutoff greater than $\sim 0.1$\,GeV.
\end{abstract}

\end{titlepage}

\setcounter{page}{2}

\tableofcontents

%\newpage

\section{Introduction}

There is increasing evidence that many effective field theories do not possess a sensible UV completion into a quantum theory of gravity. In the context of string theory, such effective theories are said to be part of the ``swampland" of theories and do not arise in any 4-dimensional compactification \cite{Vafa:2005ui}. A useful way of delineating the space of inconsistent theories is in the form of so-called ``swampland conjectures". For example, the ``distance conjecture" holds that scalar fields cannot exhibit field excursions much larger than the Planck scale \cite{Ooguri:2006in}, while the ``weak gravity conjecture" states that the lightest particles of a theory cannot carry a mass larger than their charge in Planck units \cite{ArkaniHamed:2006dz}. 

An issue of fundamental importance is whether effective theories that exhibit de Sitter vacua can be embedded within quantum gravity, or whether they too are part of the swampland. This is a topic of immense importance due to observations that our universe is accelerating \cite{Riess:1998cb}. The observations are consistent with dark energy of current value $\sim 10^{-120}\mpl^4$ ($\mpl\equiv1/\sqrt{8\pi G}$ is the reduced Planck mass) and an equation of state $w\sim -1$. The simplest version of this is that of a positive cosmological constant, and so it is natural to ask if this can be incorporated within string theory. In Ref.~\cite{Maldacena:2000mw} it was shown that an entire class of simple compactifications do not possess any de Sitter vacua regardless of the details of the internal manifold. In Ref.~\cite{Hertzberg:2007wc} it was shown that not only are there no de Sitter vacua in an entire class of simple compactifications in type IIA theory, but that a highly restrictive inequality on the 4-dimensional potential $V(\phi_i)$ could be derived. Namely it was shown that
\beq
|\nabla V|\ge c\,{V\over\mpl},
\label{SC}\eeq	
everywhere in field space, where $c=\sqrt{54/13}$ in the above IIA example (here the gradient $\nabla V$ is a vector derivative in field space and the absolute value $|\nabla V|$ means a contraction with respect to the metric on field space). Many important follow up developments have appeared, strengthening the argument that it is difficult to realize de Sitter space in string theory, including Refs.~\cite{Caviezel:2008tf}. In Ref.~\cite{Obied:2018sgi} this inequality was greatly generalized and promoted to a conjecture about all string compactifications (at least in cases where the second derivative of the potential is not too negative \cite{Garg:2018reu}). The general statement is that the coefficient $c$ is some $\mathcal{O}(1)$ number, whose absolute minimum value is not currently determined. If true, this implies that ordinary models of early universe slow-roll inflation would be in the swampland. In terms of late time acceleration, this precludes a positive cosmological constant, but does not preclude some form of quintessence \cite{Agrawal:2018own}, in which acceleration is driven by a very light rolling scalar field (at least if one can construct models in which the coefficient $c$ can be made a factor of a few smaller than what appears in known compactifications). Some follow up discussion of the viability of these conjectures includes Refs.~\cite{Denef:2018etk}.

Although the idea of quintessence is a very interesting one, it appears to be problematic from a theoretical point of view. In particular, a major objection to quintessence is that invoking it does not solve the cosmological constant problem, as it does not prevent large contributions to the cosmological constant from loop effects, Higgs potentials, QCD condensates, etc, which are generically much larger that the observed dark energy density.  Additionally, it appears to exacerbate the problem by requiring that the field is slowly rolling.  In short, this seems to replace a single fine-tuning by (at least) two; namely both $V\sim 10^{-120}\mpl^4$ and $|\nabla V|\sim 10^{-120}\mpl^3$ need to be extremely small. However, the recent swampland conjecture suggests that this effective field theory based reasoning may be too naive. In particular, if one can naturally saturate the inequality in Eq.~(\ref{SC}) it may avoid this additional fine-tuning.

The idea that quintessence in effective field theories is fine-tuned was pointed out almost as soon as the idea was first discussed. It was first noted in Ref.~\cite{Kolda:1998wq} that a viable quintessence model requires fine-tuning of not only the potential energy, but also its derivatives as well.  In Ref.~\cite{Amendola:1999er} a coupling to matter was introduced, and the resulting cosmology was determined to place (weak) constraints on its value.  In Ref.~\cite{Brax:1999yv} Coleman-Weinberg style loop corrections were studied for some quintessence potentials, finding that quantum corrections can be small, though they only studied quintessence self-coupling and in the regime $\phi>\mpl$. Coupling to fermions was considered in Ref.~\cite{Doran:2002bc}, where it was found that loop corrections ruin the quintessence potential unless severe constraints are placed on the couplings.  These authors even comment on a special form of the coupling that evades their bounds, but do not connect it with the conformally coupled theory that we will discuss, perhaps because their retention of the quadratic divergence alters the specific requirement. Loop corrections were also shown to induce a non-minimal coupling between the quintessence field and gravity in Ref.~\cite{Garny:2006wc}.  A scenario of a quintessence potential that is protected by a shift symmetry and yet whose field values remain sub Planckian, at the expense of being mildly strongly coupled, is presented in Ref.~\cite{DAmico:2018mnx}. Other related discussions includes Refs.~\cite{Chimento:2004it}.

In fact, trying to evade this multiple tuning has been extensively considered in constructions of explicit models of quintessence.  In \cite{Choi:1999xn}, a string axion model was outlined, and the fact that the tuning must occur over an entire field range was stressed.  Severe constraints on matter couplings needed to preserve the form of the potential were placed, and it was noted that such constructions require an excessive degree of tuning on the initial conditions in order to arrange a cosmologically viable model.  A model where the quintessence potential is protected from quantum corrections by a symmetry was written down in \cite{Kaloper:2008qs}; here, the quintessence field couples to a four form, and is only derivatively coupled to matter.  A model based on monodromy was explored in \cite{Panda:2010uq}; quantum corrections due to moduli stabilization, the Kahler potential and the warped throat geometry were computed, as well as associated cosmological effects.  Again in this model, matter coupling is argued to be suppressed.  Observational constraints on these types on models are explored in more detail in \cite{Gupta:2011yj}.  Finally, a model based on supersymmetric large extra dimensions is constructed in \cite{Cicoli:2012tz}.  There, the quintessence was constructed out of a fiber modulus, which naturally suppresses coupling to matter and avoids other typical tuning problems.  

The common thread through all these models is the search for specific setups within the framework of supersymmetry where quantum corrections to the quintessence potential are absent (usually due to some symmetry).  Some suppression of the coupling to matter is essential as well, due to the large corrections these otherwise generate.  The swampland criteria are generally not adhered to in any of these constructions, which either explicitly invoke small slopes, or large field excursions, or both.  It remains to be shown whether any model that satisfies the swampland bounds can also alleviate the double tuning inherent in generic quintessence models, and indeed whether such a model can be self-consistently constructed.

In this paper, we systematically study the fine-tuning in quintessence models, paying attention to the inequality in Eq.~(\ref{SC}) in the context of string compactifications. Our goal is to examine whether this inequality can be naturally saturated, hence alleviating multiple fine-tunings, as has been claimed. Firstly, we begin by discussing simple compactifications in string theory, these being the best examples of models which do indeed lead to $V$ being related to $|\nabla V|$, and whose effective field theory is under some level of control. These have energy densities that are dominated by huge classical effects, preventing them from being realistic dark energy models, but they are taken as useful inspiration for what a dark energy model may qualitatively look like. We then turn to our primary investigation of quantum effects in generic theories, whose typical value is much larger than the observed dark energy density. We search for special classes of quintessence theories that can preserve the relation between the derivative of the potential to its current value in the face of quantum corrections.  In these cases, the multiplicity of fine-tunings is an illusion and the level of fine-tuning of quintessence is the same as that of the cosmological constant. In fact we identify such a class of theories, namely those that involve conformally coupled scalars. These are stable under radiative corrections; at least when only matter particles run in the loop. However, they tend to be in conflict with observational bounds on fifth forces. Any deviation from the special form we consider, such as only coupling to dark matter, will decouple the potential and its derivative and reintroduce the second tuning. A possible loop-hole to this argument is a dark sector that is an exact copy of the Standard Model. However, when quintessons and/or gravitons run in the loop, the relation between $V$ and $|\nabla V|$ is once again spoiled, and we compute the size of this effect.  We conclude by noting that there is no evidence that a theory which can satisfy the currently formulated swampland bounds, possess the same level of fine tuning as the cosmological constant, and obey all experimental bounds, exists.

Our paper is organized as follows:
In Section \ref{SimpleClassicalExamples} we present some classical potential functions from string theory.
In Section \ref{QuantumCorrections} we compute some leading order quantum corrections.
In Section \ref{ConformalCoupling} we examine a conformally coupled scalar, and in Section \ref{DarkCopyoftheStandardModel} we discuss the possibility of a dark copy of the Standard Model.
In Section \ref{ScalarsandGravitonsintheLoop} we compute quantum corrections from quintessons and gravitons, and in Section \ref{Discussion} we discuss our results and future directions.

\section{Simple Classical Examples}\label{SimpleClassicalExamples}

In this section we provide some concrete examples of (relatively) simple compactifications in string theory, in which there exist regimes in field space where $|\nabla V|\sim V/\mpl$ arises very naturally. These will be entirely classical treatments. In the later sections we will consider quantum loop corrections from matter particles, such as the Standard Model, and analyze how they alter this conclusion.

\subsection{M-Theory}

Let us illustrate this idea in a simple compactification in M-theory, including flux on a compact curved space. Let $\phi$ be the (canonically normalized) volume modulus and let $\psi_i$ be the residual moduli. Building on earlier work in Ref.~\cite{Maldacena:2000mw}, it was shown in Ref.~\cite{Obied:2018sgi} that the 4-dimensional potential takes the following simple form
\beq
V = A_G(\psi_i)\,e^{-{10\over\sqrt{14}}\phi/\mpl} + A_C(\psi_i)\,e^{-{6\over\sqrt{14}}\phi/\mpl},
\eeq
where $A_G$ is non-negative and $A_C$ can have either sign. By taking a derivative with respect to $\phi$ it is simple to show that $|\nabla V| \ge c\,V/\mpl$ everywhere in field space, with $c=6/\sqrt{14}$ for $A_C>0$ and $c=10/\sqrt{14}$ for $A_C\leq0$. Furthermore, this inequality can be {\em saturated} by placing the matter fields $\psi_i$ at some extrema, $\partial V/\partial\psi_i = 0$, and going to large positive or negative $\phi$ as appropriate. Therefore, by allowing the heavy fields in the theory $\psi_i$ to all relax to their local minima, one finds that the inequality is saturated. Thus, as long as the overall scale of $V$ is very small (which itself appears to involve fine-tuning), $\phi$ could act as a potential candidate for quintessence.

\subsection{Type IIA Theory}

As a slightly more complicated example, consider type IIA string theory. Following Ref.~\cite{DeWolfe:2005uu}, we consider the supergravity limit, compactified on a Calabi-Yau, allowing for several $p$-form fields ($p=0,2,4$) as well as D6-branes and O6-planes. This leads to interesting models that can essentially stabilize all moduli in a regime of parametric control (though the validity of the full description of these solutions is called into question in \cite{Acharya:2006ne}). In Ref.~\cite{Hertzberg:2007wc} (also see Ref.~\cite{Hertzberg:2007ke}) it was shown that the general form of the classical 4-dimensional potential $V$ can be expressed in terms of (the canonically normalized) dilaton $\phi_d$ and the volume modulus $\phi_v$ as follows
\beq
V = A_3(\psi_i)\,e^{-(\sqrt{6}\,\phi_v+\sqrt{2}\,\phi_d)/\mpl}+\sum_p A_p(\psi_i)\,e^{-(\sqrt{2\over3}(p-3)\phi_v+2\sqrt{2}\,\phi_d)/\mpl}+A_6(\psi_i)\,e^{-{3\over\sqrt{2}}\phi_d/\mpl},
\eeq
where $A_3,\,A_p$ are non-negative and $A_6$ can have either sign. By taking derivatives with respect to $\phi_d$ and $\phi_v$ it can be readily shown that there are no de Sitter vacua and that in fact $|\nabla V| \ge c\,V/\mpl$, with $c=\sqrt{54/13}$ \cite{Hertzberg:2007wc}. Furthermore, by going to limiting values of $\phi_v$ or $\phi_d$ one finds regions in which the potential is positive and that $|\nabla V|\sim c\,V/\mpl$ is allowed, meaning that the dilaton or the volume modulus could potentially be candidates for quintessence (although $c=\sqrt{54/13}$ is a factor of a few too large for a viable quintessence model).

\section{Quantum Corrections}\label{QuantumCorrections}

In the above we gave two examples of classical string compactifications, which not only obey the swampland conjecture $|\nabla V| \ge c\,V/\mpl$, but can saturate the bound quite naturally by letting heavy fields relax to their minima, while leaving other fields (such as the volume modulus/dilaton) free to roll. In order for these remaining fields to act as a form of quintessence in the universe today, one must ultimately construct models in which the overall vacuum energy density is extremely small $V\sim 10^{-120}\mpl^4$. For such extremely small energy densities, we are not free to compute only the leading order contributions from classical compactifications, but must also include all sorts of effects, such as those from quantum loops of particles within the Standard Model. In this section we illustrate the problems this can lead to.

\subsection{Leading Quantum Contribution}

For a simple representation of a matter sector, consider a set of massive scalars $\sigma_i$, that are coupled to the quintessence field $\phi$ and minimally coupled to gravity as follows
\beq
S = \int d^4x\sqrt{-g}\left({1\over2}(\partial\phi)^2+\sum_i\left[{1\over 2}(\partial\sigma_i)^2-{1\over2}f_i(\phi)m_i^2\sigma_i^2\right]\right),
\eeq
where the function $f_i(\phi)$ may be an exponential $f_i(\phi)=e^{c_i\phi/\mpl}$, though not necessarily so. For definiteness let us expand around $\phi=0$ and parameterize the linear piece of $f_i$ as
\beq
f_i(\phi)=1+{c_i\,\phi\over\mpl}+\ldots.
\eeq

\begin{centering}
	\begin{figure*}[t]
		\centering
		\includegraphics[height=3.7cm]{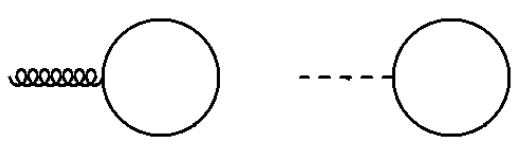}
		\caption{One-loop contributions to $V_{\rm eff}(\phi)$ from matter particles (solid lines) in the loop: left diagram is cosmological constant and right diagram is linear term in $\phi$.}
		\label{vprime}
	\end{figure*}
\end{centering}

We first compute the contribution to the cosmological constant $\Lambda$ from a scalar $\sigma_i$ running in a loop; see the diagram in the left of Figure \ref{vprime}. By expanding the metric as $g_{\mu\nu}=\eta_{\mu\nu}+h_{\mu\nu}$, the relevant 3-point interaction is
\beq
\Delta \mathcal{L} =-{1\over2} h^{\mu\nu}\sum_i\left[\partial_\mu\sigma_i\partial_\nu\sigma_i-{1\over2}\eta_{\mu\nu}(\partial\sigma_i)^2+{1\over2}\eta_{\mu\nu}m_i^2\sigma_i^2\right].
\eeq
The generated cosmological constant is the counter-term $-h^{\mu\nu}\eta_{\mu\nu}\Lambda/2$. This is readily determined to be
\beq
\Lambda = \sum_i\int \! {d^4p\over(2\pi)^4}\frac{{1\over4}p^2+{1\over2}m_i^2}{p^2+m_i^2} = \sum_i{m_i^4\over64\pi^2}\ln\!\left(\frac{m_i^2}{\mu^2}\right),
\eeq
where we have Euclideanized the integral. Here we have ignored power law divergences, as they can be re-absorbed into bare couplings, and kept only the logarithmic divergence with cut-off $\mu$; this piece is formally related to a type of RG flow of the cosmological constant.

We now consider the contribution to the effective potential from the 3-point interaction in the above Lagrangian $\Delta\mathcal L=-\sum_i c_i\,\phi\, m_i^2 \sigma_i^2/(2\mpl)$. The required counter-term at one-loop is given by the tadpole diagram in the right of Figure \ref{vprime}, which is
\beq
\Delta V_1= \sum_i{1\over2}{c_i\,\phi\,m_i^2\over\mpl}\!\int \! {d^4p\over(2\pi)^4}\frac{1}{p^2+m_i^2} = \sum_i 2{c_i\,\phi\over\mpl} {m_i^4\over64\pi^2}\ln\!\left(\frac{m_i^2}{\mu^2}\right),
\eeq
where again we have Euclideanized and only extracted the logarithmic piece of the integral.

By combining the above classical potential, generated cosmological constant, and generated linear term in $\phi$, we obtain the following effective potential $\Veff$ at one-loop
\beq
\Veff(\phi,\sigma_i)=\sum_i\left[{1\over2}m_i^2\,\sigma_i^2\left(1+{c_i\,\phi\over\mpl}\right)+{m_i^4\over64\pi^2}\ln\!\left(m_i^2\over\mu^2\right)\left(1+2{c_i\,\phi\over\mpl}\right)\right].
\eeq
Note that for generic values of $c_i$ the potential does not factorize and therefore it is not ordinarily the case that $\partial V/\partial\phi$ is related to $V$. So in general we need
\beq
c_1=c_2=\ldots=c
\eeq
in order to make this relation natural. However, even assuming this, we see that the factor of 2 in the final term still prevents factorization.

\subsection{More General Corrections}

More generally, suppose the Lagrangian has a (classical) potential term $V$ of the form that factorizes
\beq
V(\phi,\sigma_i) = f(\phi)\sum_i A_i(\sigma_i).
\eeq
Then, using the Coleman-Weinberg formula for the one-loop effective potential, and only allowing $\sigma_i$ particles to run in the loop (we will consider $\phi$ and/or gravitons in the loop in Section \ref{ScalarsandGravitonsintheLoop}), we obtain
\beq
\Veff(\phi,\sigma_i)=f(\phi)\sum_i A_i(\sigma_i) + f(\phi)^2\sum_i{(A_i''(\sigma_i))^2\over 64\pi^2}\ln\!\left(A_i''\over\mu^2\right).
\eeq
Hence, the factorization of the initial classical potential is broken in the quantum corrected potential for any non-trivial $f(\phi)$. So even if $f(\phi)=e^{c\,\phi/\mpl}$, or similar, which means that the classical potential can obtain $|\nabla V|=c\,V/\mpl$, with $\sigma_i$ set to their equilibrium values, this is ruined after quantum corrections.

It is possible that both $V$ and $\partial V/\partial\phi$ are still small enough for $\phi$ to act as a viable quintessence field, but here it represents (at least) a double fine-tuning, as any relation that may have been imposed on the tree level potential is spoiled by quantum effects. Note that this applies through phase transitions, so that if at a certain time the potential and its derivative are tuned to be small, this will generically no longer hold after the phase transition.

\section{Conformal Coupling}\label{ConformalCoupling}

In the above we found that quantum corrections from matter loops generically spoil any relationship between $V$ and $\nabla V$ that may have been imposed on the classical theory. However one can investigate if there is any type of special class of coupling that can avoid this problem. The new ingredient that could lead to this possibility is to consider a quintessence field $\phi$ that not only couples to the mass term of matter fields, but also couples to their kinetic terms as well. At the level of effective theories, this is a perfectly reasonable possibility. 

\subsection{One-Loop Analysis}\label{OneLoopAnalysis}

For the sake of greater generality, we will couple $\phi$ both to a set of scalars $\sigma_i$ as well as fermions $\psi_j$ as follows
\beq
\mathcal L = {1\over2}(\partial\phi)^2+\sum_i \left[{1\over2}g_i(\phi)(\partial\sigma_i)^2-\frac12f_i(\phi)m_i^2\sigma_i^2\right]+\sum_j\left[i\,h_j(\phi)\bar\psi_j\slash\!\!\!\partial\psi_j - j_j(\phi)m_j\bar\psi_j\psi_j\right],
\label{scalarfermion}\eeq
where we have allowed various coupling functions: $g_i,\,f_i$ to scalars and $h_j,\,j_j$ to fermions. 

Again, we will focus on one-loop quantum corrections from matter particles running in the loop. This can be computed directly by expanding the functions $g_i,\,f_i,\,h_j,\,j_j$ around some reference $\phi$ value (say $\phi=0$) and proceeding perturbatively. While this is straightforward, it is rather tedious for non-trivial functions. Instead we can compute the effects of matter loops very easily, by noting that the external $\phi$ is slowly varying. In this case it is convenient to re-scale the matter fields as follows
\beq
\sigma_i\to\sigma_i/\sqrt{g_i(\phi)},\,\,\,\,\,\,\,\psi_j\to\psi_j/\sqrt{h_j(\phi)},
\eeq
which renders the kinetic terms for the matter sector canonical for slowly varying $\phi$. Then one can readily use the Coleman-Weinberg formula for the one-loop effective potential. We find
\bea
V_{\rm eff}(\phi,\sigma_i,\psi_i)=\sum_i\left[\frac12f_i(\phi)m_i^2\sigma_i^2 +\frac{f_i(\phi)^2}{g_i(\phi)^2}{m_i^4\over64\pi^2}\ln\!\left(\frac{m_i^2}{\mu^2}\right)\right]\nonumber\\
+\sum_j\left[j_j(\phi)m_j\bar\psi_j\psi_j-\frac{j_j(\phi)^4}{h_j(\phi)^4}{m_j^4\over64\pi^2}\ln\!\left(\frac{m_j^2}{\mu^2}\right)\right].\!\!\!\!\!
\eea
In order for the original form to be preserved without additional tuning, it is necessary that all the functions are related to a single function $f(\phi)$ as follows
\beq
f(\phi)=f_i(\phi)=g_i(\phi)^2=j_j(\phi)=h_j(\phi)^{4/3}.
\label{conformalrelations}\eeq
These conditions are precisely those for the quintessence field to be {\em conformally} coupled to matter. This ensures that the coupling completely factorizes for any choices of mass parameters for scalars or fermions.  This is reminiscent of the observation in \cite{Hui:2010dn} that the enhanced symmetry of this theory prevents corrections to the equivalence principle from appearing due to matter loops.

\begin{centering}
	\begin{figure*}[t]
		\centering
		\includegraphics[width=\textwidth]{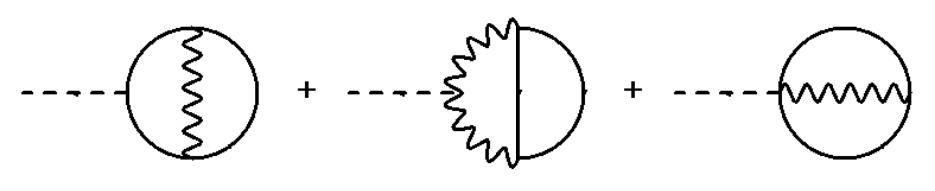}
		\caption{Some representative contributions to $V_{\rm eff}(\phi)$ allowing matter particles (solid lines) and gauge bosons (wiggly lines) in the loop.}
		\label{full}
	\end{figure*}
\end{centering}

\subsection{All Orders Analysis}\label{AllOrdersAnalysis}

The above derivation assumed that there were no interactions in the matter sector, so that the analysis could be truncated at one-loop. However, it is very important to consider interactions in the matter sector, such as from gauge interactions. For example, at two-loops one has the diagrams of Figure \ref{full}. In this case a direct computation becomes more and more difficult. However for the case of a conformally coupled scalar there is a straightforward way to proceed.

Consider the following action for the matter sector
\beq
S_M = \int d^4x\sqrt{-g}\,f(\phi)\,\mathcal{L}_M(\sigma_i,\psi_j,A^\mu_k,g_{\mu\nu}\sqrt{f(\phi)}),
\eeq
where we have indicated that this sector may contain scalars $\sigma_i$, fermions $\psi_j$, gauge bosons $A^\mu_k$, etc. This action defines a conformally coupled scalar (in the Einstein frame). Namely, the coupling is of the following simple form: one merely replaces $g_{\mu\nu}\to g_{\mu\nu}\sqrt{f(\phi)}$ for a set of matter fields that are otherwise minimally coupled to gravity. If the matter sector involves only massive particles and no interactions this will reproduce the above matter Lagrangian in Eq.~(\ref{scalarfermion}), with relations given in Eq.~(\ref{conformalrelations}). In the case of interacting particles, this provides a complete non-linear definition of what we mean by {\em conformal coupling} (note we are {\em not} assuming the matter sector carries {\em conformal symmetry}, which is a separate idea).

As above, we now wish to consider quantum effects from matter particles running in loops, but not scalars or gravitons (we turn to this issue in Section \ref{ScalarsandGravitonsintheLoop}). Since $\phi$ is therefore only external and slowly varying, it is convenient to define an auxiliary metric $G_{\mu\nu}$ as
\beq
G_{\mu\nu}\equiv g_{\mu\nu}\sqrt{f(\phi)}.
\eeq
Then, the quantization of matter species $\sigma_i,\,\psi_j,\,A^\mu_k$ proceeds in the standard way as for any matter fields in a background curved space-time $G_{\mu\nu}$. We know that, in general relativity, the leading order interactions have a universal coupling to the graviton, and that this maintained under renormalization (otherwise there would be a gauge anomaly as it would destroy the diffeomorphism redundancy needed to consistently couple to a massless spin 2 field). Hence, since the matter sector is also assumed to universally couple to $\phi$, that universal coupling will also be maintained under renormalization. 

In flat space, due to any number of loops in the matter sector, one expects a potential function, $A_{\rm eff}(\sigma_i,\psi_j)$, for the matter fields to be generated. In the presence of a background metric $G_{\mu\nu}$, this provides a contribution to the action of the form $\Delta S = -\int d^4x\sqrt{-G}\,A_{\rm eff}(\sigma_i,\psi_j)$. By making $\phi$ explicit, we can read off the full effective potential as
\beq
V_{\rm eff}(\psi,\sigma_i,\psi_j) = f(\phi)\,A_{\rm eff}(\sigma_i,\psi_j).
\eeq
Hence, the factorization is maintained exactly under any number of matter loops, including interactions. This suggests that a conformally coupled scalar may be a very natural candidate for quintessence. In particular, if $f(\phi)$ has the property
\beq
f'(\phi)\sim{f(\phi)\over\mpl},
\eeq
(which is quite reasonable for some of the moduli in string compactifications) then not only can this saturate the swampland bound of Eq.~(\ref{SC}), but it can lead to a potentially viable model of dark energy. At this stage this only involves a {\em single} fine-tuning, namely to explain why $A_{\rm eff}(\sigma_i,\psi_j)$ is very small (at least when matter fields are placed at their vacuum expectation values).

\subsection{Fifth Forces}\label{FifthForces}

At first sight the above scenario of a conformally coupled quintessence field seems rather promising. However, it has immediate observational consequences. In particular, it will couple to the Standard Model of particle physics, and so it is subject to fifth force constraints. In the case of conformal coupling, the field couples to matter universally, and so it does not upset tests of the weak equivalence principle (while a non-conformally coupled scalar would violate the weak equivalence principle, leading to the bound $c\lesssim 10^{-5}$ \cite{Touboul:2017grn}). Nevertheless, since a quintessence field should have an extremely small mass (typically $m_\phi\sim H_0\sim 10^{-33}$\,eV), it is subject to solar system tests of gravity, and affects light bending. In this case it can be constrained in a fashion similar to Brans-Dicke models, which leads to the bound $c\lesssim 0.01$ \cite{Bertotti:2003rm}. 

At this point one may conclude that fifth forces may be generic predictions of quantum theories of gravity that obey the conjecture of Eq.~(\ref{SC}) and incorporate dark energy. Arguably, the idea of trying to find models that simultaneously obey the conjecture as well as current observational bounds may be on the wrong track, as the generic predictions of this framework seem to have already been invalidated. Nevertheless there may be possible ways to avoid this conclusion, which we will discuss in the remainder of this paper.

\section{Dark Copy of the Standard Model}\label{DarkCopyoftheStandardModel}

We now discuss what appears to be the only way to avoid the above obstructions from multiple fine-tunings and fifth force constraints. Consider a theory in which the quintessence field $\phi$ is conformally coupled to a dark sector, but not directly coupled to the Standard Model. To be precise, we consider the action
\beq
S = \int d^4x\sqrt{-g}\left[\,f(\phi)\,\mathcal{L}_{DS}(\tilde\sigma_i,\tilde\psi_j,\tilde{A}^\mu_k,g_{\mu\nu}\sqrt{f(\phi)})+\mathcal{L}_{SM}(\sigma_i,\psi_j,A^\mu_k,g_{\mu\nu})\right],
\eeq
where $\mathcal{L}_{DS}$ is the dark sector Lagrangian and $\mathcal{L}_{SM}$ is the Standard Model Lagrangian. This will trivially evade both laboratory and solar system based fifth force constraints. However, if this dark sector involves the dark matter of the universe, then there are still non-trivial constraints on the equivalence principle on galactic scales \cite{Kesden:2006vz}.

On the other hand, the model appears to re-introduce the fine-tuning problems that we discussed in Section \ref{QuantumCorrections}, since we are not conformally coupled to the entire matter sector. In particular, the contributions to $V$ from Standard Model particles destroys the relation between the slope and the potential. If we denote the effective potential in the dark sector as $A_{eff,DS}$ and the effective potential in the Standard Model sector as $A_{eff,SM}$, then the total effective potential is
\beq
V_{\rm eff}(\phi,\sigma_i,\tilde\sigma_i,\psi_j,\tilde\psi_j)=f(\phi)\,A_{eff,DS}(\tilde\sigma_i,\tilde\psi_j)+A_{eff,SM}(\sigma_i,\psi_j),
\eeq
which shows that $V$ and $\partial V/\partial\phi$ are generically unrelated to each other.

However, there appears to be one way to avoid this problem. Suppose the dark sector is an exact copy of the Standard Model, including the same particle content, the same couplings, etc. (Note that we are not postulating this for the dark {\em matter} necessarily, only the dark {\em sector}. We can imagine that such a dark sector is not as populated as the visible sector in the early universe, hence avoiding problems with big bang nucleosynthesis and other cosmological constraints~\cite{Kolb:1985bf}. Also, we can be agnostic about the details of the dark matter; for example, it could be of the ``mirror dark matter" variety \cite{Foot:1991bp} or it could simply be primordial black holes, etc). In this special case, we have $A_{eff,DS}=A_{eff,SM}$. By placing matter fields at their vacuum expectation values, we then have
\beq
{|\nabla V|\over V}={f'(\phi)\over f(\phi)+1}.
\eeq
Therefore, for simple choices of the function $f(\phi)$ we could naturally have $|\nabla V|\sim V/\mpl$, while evading fifth forces in the visible sector entirely. In addition, we can trivially extend this to multiple dark copies of the Standard Model if desired.

Note that even if we have a dark copy of the Standard model, there are two possible dimension 4 operators allowed that can couple the two sectors
\beq
\Delta\mathcal{L} = -{\epsilon\over4} F_{\mu\nu}\tilde F^{\mu\nu}-\delta\,H^\dagger H \tilde H^\dagger \tilde H,
\eeq
where ($\tilde F$) $F$ is the (dark) hypercharge field strength and ($\tilde H$) $H$ is the (dark) Higgs field. In order to not spoil the above result, we would need these two sectors to remain essentially decoupled (except obviously through gravity; more on this in the next Section). This means that the coefficients $\epsilon$ and $\delta$ must be extremely small. We note that such an assumption is stable under renormalization, so although it may appear to be a type of fine-tuning, it is a technically natural assumption.

\section{Quintessons and Gravitons in the Loop}\label{ScalarsandGravitonsintheLoop}

Thus far, we have shown that so long as we only consider matter particles in the loop, and so long as they are conformally coupled, we can maintain the relation $|\nabla V|\sim V/\mpl$ under renormalization due to the non-perturbative proof of Section \ref{AllOrdersAnalysis}. However, that proof relied on taking $\phi$ and the metric as external fields only. It does not hold if we consider a member of the gravitational sector -- either the quintesson or the graviton -- to run in the loop. 

\begin{centering}
	\begin{figure*}[t]
		\centering
		\includegraphics[height=4.5cm]{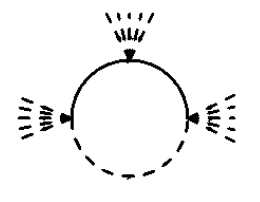}
		\caption{One-loop correction to $V_{\rm eff}(\phi)$ including a $\phi$ (dashed line) running in the loop with a matter particle (solid line).  The spray of external lines is shorthand for the full sum of diagrams with any number of external $\phi$ particles.}
		\label{1loopphi}
	\end{figure*}
\end{centering}

If the matter sector begins at quadratic order in the fields, then the first diagram that can involve $\phi$ or gravitons in the loop begins at two-loop order, which is moderately complicated. Therefore, for the sake of simplicity, let us consider a matter sector that includes a term that has a {\em linear} potential for the matter fields $V(\sigma)\propto\sigma$, which leads to the possibility of $\phi$ running in the loop at only one-loop order. For concreteness, consider the following Lagrangian with scalars $\sigma_i$ and a conformally coupled scalar $\phi$
\beq
\mathcal{L} = {1\over2}(\partial\phi)^2+\sum_i\left[{1\over2}\sqrt{f(\phi)}\,(\partial\sigma_i)^2 - f(\phi)\,\gam_i^3\,\sigma_i\right],
\eeq
where $\gam_i$ is a mass scale. This may seem to be an unusual action, since we are not expanding around a vacuum, but it will suffice to illustrate the main point. In addition to a tiny renormalization of $\gamma_i$ and the generation of a tiny mass for $\sigma_i$, there is a potential for $\phi$ generated from the diagrams of Figure \ref{1loopphi}. Focusing on this new contribution, we find the effective potential is
\beq
V(\phi,\sigma_i) = \sum_i\left[ f(\phi)\,\gam_i^3\,\sigma_i + {9\,\gam_i^6\,f'(\phi)^2\over 512\pi^2\sqrt{f(\phi)}}\ln\!\left(M_i^2\over\mu^2\right)\right],
\eeq
where $M_i^2$ is some effective mass associated with the potential, whose details are not too important. We see that for generic choices of the conformal coupling function $f(\phi)$ this evidently spoils the relation between $V'$ and $V$, since the $\phi$ dependence does not factorize. In order to obtain factorization, we would need $f'(\phi)^2/\sqrt{f(\phi)}\propto f(\phi)$, which implies $f(\phi)=(1+c\,\phi/\mpl)^4$. However this special form for $f(\phi)$ is peculiar to the assumption of a linear potential in $\sigma$ and does not work for generic matter.

\begin{centering}
	\begin{figure*}[t]
		\centering
		\includegraphics[height=3.3cm]{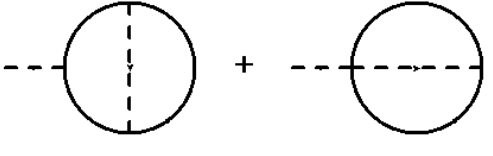}
		\caption{Two-loop correction to $V_{\rm eff}(\phi)$ including a $\phi$ running in the loop.}
		\label{2loopphi}
	\end{figure*}
\end{centering}

Furthermore, for generic coupling function $f(\phi)$ we can Taylor expand the above to obtain a mass for the scalar as $\Delta m_\phi\sim c^2\,\gam^3/(4\pi\mpl^2)$. A related issue occurs when we consider massive fields $\psi_i$. At two-loops we have diagrams such as those in Figure \ref{2loopphi}, which again spoil the relationship between $V'$ and $V$. Furthermore, they generate a mass for the quintessence field as
\beq
\Delta m_\phi\sim {c^2\,m_\psi^3\over(4\pi)^2\mpl^2}\,.
\eeq 
There are similar corrections from gravitons running in a loop, although we will not go into those details other than to note that they introduce $\Delta m_\phi\sim\gam^3/(4\pi\mpl^2)$ or $\Delta m_\phi\sim m_\psi^3/((4\pi)^2\mpl^2)$ for the above models.

Assuming $c\sim 1$, the quintesson and graviton loop contributions to the mass of $\phi$ are of a similar order. Therefore, barring some unexpected cancellations (see discussion), we anticipate that the scalar mass will obtain corrections $\Delta m_\phi\sim m_\psi^3/((4\pi)^2\mpl^2)$. This is suppressed by two powers of the Planck mass, rendering it rather small. This should be compared to a non-conformally coupled scalar, for which the mass correction from just a matter particle in the loop gives $\Delta m_\phi\sim m_\psi^2/(4\pi\mpl)$, which is suppressed by only one power of the Planck mass. The conformal coupling leaves only this quintesson/graviton loop correction, which is therefore much smaller. 

However this contribution is not negligible. In particular, if the matter particles include the entire (dark) Standard Model, including the (dark) top quark, then this leads to a mass for the quintessence field of the order $\Delta m_\phi\sim 10^{-24}$\,eV. While this is very small compared to masses of particles in the Standard Model, it is still much larger than the observed Hubble value $H_0\sim 10^{-33}$\,eV, which is the natural value for quintessence. Intriguingly, if the heaviest particle the quintessence couples to has mass $m_\psi\sim 0.1$\,GeV, such as pions, we obtain $\Delta m_\phi\sim H_0$ the desired mass value. However, only coupling to a portion of the full sector compromises the cancellation of all matter loops.

\section{Discussion}\label{Discussion}

We have identified two obstacles to implementing string theory inspired quintessence models; (i) to avoid multiple fine-tuning one would like to conformally couple the quintessence field, which generically leads to fifth forces; if this can be avoided by only coupling to the dark sector in a special fashion (a dark copy of the Standard Model) then (ii) quintessons and graviton loop corrections nevertheless introduce some amount of fine-tuning from matter particles with masses above $\sim 0.1$\,GeV.

A possible way out of these conclusions is to suppose that the effective theory for the quintessence field is only valid up to $\Lambda_{UV}\sim 0.1$\,GeV (or smaller), forbidding heavier particles, such as (dark) Higgs or tops, for example, from running in the loop. This begs the question as to what might be the new physics entering at this scale, which is an interesting avenue for exploration. 

Another question is if a cancellation can arise between quintessons and gravitons from a special choice of the coupling function $f(\phi)$. We do not have evidence that such a cancellation can occur for a non-trivial matter sector, nor do we have evidence that it could persist to higher loop order, but it may be worth exploring carefully.

A possible way to avoid fifth forces is to appeal to a type of screening mechanism (for a review see Ref.~\cite{Joyce:2014kja}). The most relevant one is the ``chameleon mechanism" \cite{Khoury:2003rn}, whereby the effective mass of the quintessence field becomes large when the local density is large, and thereby shuts down the extra scalar force. Such models appeal to an unusual diverging potential function and are known to exhibit their own levels of fine-tuning. Rather stringent constraints already exist on these models \cite{Burrage:2016bwy}, but it is worth exploring if they can be realized within string theory.

Generically, however, it appears that quintessence models face significant fine-tuning and/or fifth forces beyond the usual problems for pure vacuum energy. We note that this does not necessarily mean that the swampland conjectures are wrong. It could be that they are correct, but then obtaining a natural and consistent model for dark energy in the framework of string theory would appear rather difficult. Alternatively, such conjectures may need further refinement. These important issues deserve further investigation.

\section*{Acknowledgments}

We would like to thank Jose Blanco-Pillado, Robert Foot, Jonathan Heckman, Justin Khoury, Matt Kleban, Jeremy Sakstein, Adam Solomon, and Cumrun Vafa for helpful discussions. The work of MPH and MS is supported in part by National Science Foundation grant PHY-1720322. The work of MT is supported in part by US Department of Energy (HEP) Award DE-SC0013528, and by NASA ATP grant NNH17ZDA001N.

%\newpage

\end{document}